# Atomic and Electronic Structure of Si Dangling Bonds in Quasi-free Standing Monolayer Graphene

Yuya Murata,[1] Tommaso Cavallucci,[1] Valentina Tozzini,[1] Niko Pavliček,[2] Leo Gross,[2] Gerhard Meyer,[2] Makoto Takamura,[3] Hiroki Hibino,[3,*] Fabio Beltram,[1] and Stefan Heun[1,†]

[1]*NEST, Istituto Nanoscienze-CNR and Scuola Normale Superiore, Piazza San Silvestro 12, 56127 Pisa, Italy*

[2]*IBM Research-Zurich, Säumerstrasse 4, 8803 Rüschlikon, Switzerland*

[3]*NTT Basic Research Laboratories, 3-1 Morinosato Wakamiya, Atsugi, Kanagawa 243-0198, Japan*

ABSTRACT

Si dangling bonds without H termination at the interface of quasi-free standing monolayer graphene (QFMLG) are known scattering centers that can severely affect carrier mobility. In this report, we study the atomic and electronic structure of Si dangling bonds in QFMLG using low-temperature scanning tunneling microscopy/spectroscopy (STM/STS), atomic force microscopy (AFM), and density functional theory (DFT) calculations. Two types of defects with different contrast were observed on a flat terrace by STM and AFM. Their STM contrast varies with bias voltage. In STS, they showed characteristic peaks at different energies, 1.1 and 1.4 eV. Comparison with DFT calculations indicates that they correspond to clusters of 3 and 4 Si dangling bonds, respectively. The relevance of these results for the optimization of graphene synthesis is discussed.

---

[*] present address: Kwansei Gakuin University, 2-1 Gakuen, Sanda, Hyogo 669-1337, Japan
[†] corresponding author. e-mail: stefan.heun@nano.cnr.it

Graphene, a two-dimensional sheet of carbon atoms, is attracting much interest in high-speed electronics applications owing to its high carrier mobility.[1] The latter, however, can be severely reduced by scattering centers introduced by its supporting substrate, and this stimulates investigations for the establishment of synthesis methods that minimize graphene interaction with the substrate. Graphitization of a silicon carbide (SiC) surface is promising for electronic applications, since graphene is grown directly on an insulating substrate with no transfer process. As a first step of graphene synthesis on a SiC(0001) surface, a carbon layer, which is called buffer layer, is formed by annealing the SiC substrate in vacuum or an inert-gas atmosphere for sublimation of surface Si atoms. Though the buffer layer consists of a honeycomb structure of carbon atoms, it does not have the electronic properties of graphene, due to formation of covalent bonds to the substrate.[2] The buffer layer can be detached from the substrate and transformed to $sp^2$ hybridized carbon (graphene), either by further thermal decomposition of the SiC underneath the buffer layer, or by intercalation of H atoms at the interface between buffer layer and substrate.[3,4] The former is called epitaxial monolayer graphene (EMLG) and the latter quasi-free standing monolayer graphene (QFMLG). The carrier mobility of QFMLG shows a weaker temperature dependence than EMLG.[5] This was attributed to a reduced interaction between carriers in QFMLG and substrate phonons. However, the mobility of QFMLG (~6600 $cm^2V^{-1}s^{-1}$)[6] is lower than in free-standing graphene, thus indicating the existence of remaining carrier-scattering centers.

Based on photoelectron spectroscopy measurements, it was suggested that Si dangling bonds without H termination at the interface donate charge to graphene and act as charged scattering centers.[7] In a previous report, we studied the correlation between the morphology of QFMLG and $T_H$, the substrate temperature during H intercalation.[8] In scanning tunneling microscopy (STM) images measured at room temperature on QFMLG samples formed at $T_H$ = 600 and 800 °C, depressions with a width of 1 nm were observed on flat terraces. These depressions distribute with the periodicity of the SiC(0001) quasi-(6×6) reconstruction, i.e., with the periodicity of the buffer layer. This suggests that these depressions correspond to the positions of Si dangling bonds at the interface, since it was theoretically reported that there is a variation of 1 eV in the hydrogenation energy of Si sites within the unit cell of a buffer layer model.[9] This strong preference of H to bind in symmetric locations might in certain conditions determine also the symmetric location of the last non-hydrogenated sites, i.e. possibly remaining H vacancies. It was also found that the density of these depressions on a QFMLG sample formed at $T_H$ = 1000 °C is two orders of magnitude smaller than that on samples formed at $T_H$ = 600 and 800 °C.[8] As $T_H$ increases, the dissociation of $H_2$ molecules, the intercalation of H atoms, and their diffusion along the graphene-substrate interface are promoted, and this leads to a better H intercalation, i.e., less Si dangling-bond formation.[10] A similar distribution of Si dangling bonds at the interface of quasi-free standing bilayer graphene was reported by noncontact scanning nonlinear dielectric potentiometry.[11] However, the detailed structure of Si dangling bonds and the interaction with graphene are still poorly understood.

In this report, we study the atomic and electronic structure of these Si dangling bonds in QFMLG using low-temperature STM, atomic force microscopy (AFM), and density functional theory (DFT) based calculations.

The sample preparation process is identical to the one reported in Refs. [8] and [12]. In brief, a 4H-SiC(0001) substrate was used as starting point. The sample was cleaned by annealing in $H_2$ at 33 mbar and 1500 °C for 5 min. A buffer layer was formed by annealing in Ar atmosphere at 800 mbar and 1650 °C for 5 min. Finally, the sample was annealed in $H_2$ at 1013 mbar and $T_H$ = 1000 °C for 60 min for H intercalation. The sample was then mounted in an ultra-high vacuum chamber and annealed at 250 °C for 1 hour and 600 °C for 10 min. The sample was characterized by STM/AFM in a homebuilt combined STM and AFM operating under ultrahigh vacuum conditions (base pressure p < $10^{-10}$ mbar) at a temperature $T$ = 5 K. The microscope is equipped with a qPlus sensor [13,14] with an eigenfrequency of $f_0$ = 31034 Hz, a stiffness $k$ = 1800 N/m and a quality factor on the order of $10^5$. Bias voltage $V$ was applied to the sample. The AFM was operated in frequency-modulation mode[15] at an oscillation amplitude of $A$ = 50 pm. STM images were obtained in constant-current mode, in which a feedback adjusts the height of the tip above the surface so that the tunnel current is kept constant. Scanning tunneling spectroscopy (STS) was measured after stabilizing the tip-sample distances at a bias voltage $V_{stab}$ and tunneling current $I_{stab}$. AFM images were obtained in constant-height mode, in which the tip is scanned without a feedback parallel to the surface while the shift in resonance frequency, $\Delta f$, is recorded. AFM images were taken at distances for which $\Delta f$ increases (to less negative values) as the tip-sample distance decreases.

DFT Calculations were performed with the Quantum ESPRESSO[16] (QE, version 5.3.0) code, with a previously tested setup[17] using ultrasoft pseudopotentials,[18] Perdew-Burke-Ernzerhof exchange-correlation functional[19] with van der Waals D2 correction scheme.[20] The chosen calculation supercell has a $\sqrt{31} \times \sqrt{31}$ R8.95 periodicity with respect to SiC, which is a good approximation of the quasi-(6x6) symmetry, and 7×7 (R21.787) with respect to graphene. This choice has the advantage of including a relative small number of atoms in the model system (284) allowing extensive calculations, at the expense of only a 0.7° rotation of graphene with respect to SiC. In the simulations, this cell hosts a single H vacancy per unit cell, corresponding to the maximum vacancy concentration observed in experiments. Further details of the calculation setup are given in the SI.

Figures 1(a-h) show STM images taken at various bias voltages around a void in the SiC substrate with a depth of 0.25 nm. The void may be formed by etching of the SiC substrate during H intercalation at 1000 °C [8]. In the STM image of Fig. 1(a) at +1.5 V bias, two types of bright features with different contrast are observed on a flat terrace. We label the brighter features A, and the less bright ones B. These features were observed consistently in various areas of the sample. As shown in the SI, features A and B are representative for two groups of features on the sample, because the height histogram of 166 features analyzed clearly shows two peaks at 0.07 nm and at 0.13 nm, corresponding to B and A, respectively. Figures 1(a) to (h) show how the contrast of the A and B features changes with bias voltages. The void in the SiC substrate was used as a reference to locate the same area in the images with different bias voltages. Features B are clearly visible at bias +1.5 V, but their contrast becomes weak below +1.0 V. On the other hand, the A features are clearly visible in a bias range from +0.5 V to +1.5 V, while for lower bias their contrast becomes much weaker. The left edge of the void changes the contrast with different bias voltages, similar to the contrast of features A and B.

Figures 1(i) and (j) are STM and AFM images, respectively, taken at the area indicated by the square in Fig. 1(a). In the STM image of Fig. 1(i) at +0.1 V and in the AFM image of Fig. 1(j), the honeycomb lattice of graphene is clearly observed also within the two features, indicating that they are not defects of the graphene lattice itself or adsorbates on graphene. Besides, point defects in graphene would result in a clear electronic signature which is obviously not observed [REF]. The AFM image in Fig. 1(j) shows a darker contrast at the position of the A and B features, i.e. a more negative Δf at the position of the features compared to the surrounding flat area. Besides, the lateral extension of feature B is larger than that of A.

These qualitative observations can be quantified by analysis of cross-sectional profiles taken from these STM data. Figures 2(a) and (b) show line profiles of typical A and B features at various bias voltages. Their apparent height varies with bias voltage. Features A appear brighter than features B in the range of bias voltage from +1.5 V to +0.1 V, but at -0.05 V they appear darker than features B and the surrounding flat area. Figure 2(c) shows line profiles from the AFM image in Fig. 1(j) taken along the [11$\bar{2}$0] direction. They show that Δf is more negative at the position of the features compared to the surrounding flat area. Besides, at the position of feature B, both the lateral extension and the intensity of the frequency decrease are larger than those for feature A. Any decrease of the frequency corresponds to an increased tip-sample distance and, since we are operating the AFM in constant height mode, it indicates a depression of the surface morphology [22,23]. Therefore, these data imply that at the positions of the features, graphene is deformed towards the substrate, and the depression is more pronounced at the position of features B than A. The possibility that an atom or a cluster larger than H is intercalated at the position of the two features can be excluded, because in this case the graphene would be deformed towards the vacuum, not the substrate.

This picture is confirmed by DFT calculations. We evaluated the structure and electronic properties of many vacancies with different number of missing H atoms, from 1 to 13 (see the SI for details on the calculations performed). The results of this extensive sampling are reported elsewhere.[24] All of them display intrusions of the graphene sheet towards the substrate that increase with the size of the vacancy. Figure 2(d) reports the z displacement of C atoms along the [11$\bar{2}$0] direction, for three selected cases of vacancies with 1, 3 and 4 missing atoms (1H, 3H and 4H, also illustrated in Fig. 4). The 1H vacancy displays a very small inward bending of a few pm that would be barely detectable by AFM;[23] similarly, the 2H vacancies display a very small inward bending less than 5pm (not reported). 3H and 4H vacancies appear to be the only ones compatible with observations, being the smallest with a detectable bending, between 20 and 40pm, while larger vacancies display bending of more than 50-60 pm which is likely to be much larger than the observed range. Figure 4 (third row) reports the AFM-like images generated from iso-charge density surfaces. In accordance to the experimental observations, 3H and 4H images show smaller and larger contrast, respectively, while the 1H vacancy is almost undetectable. We remark that these calculations were performed in almost-neutral conditions because the DFT calculations do not reproduce properly the natural doping of the QFMLG. This is mainly due to the poor representation of the substrate, as verified elsewhere.[24] Furthermore, an artificially-induced doping by positively charging graphene might increase the inward bending because the positive charge localizes under the vacancy. However, detectable bending in 1H is observed only for doping at least three times larger than in experiment.

In order to characterize the electronic states of the two types of features, we performed STS measurements. Figures 3 (a) and (b) show the STS results on the features and from the surrounding flat area. Each curve was obtained by averaging over 20 spectra taken from different features of the same type. All three spectra show a gap with a width of approximately 0.1 V at around 0 V, and a dip at +0.25 V, see Fig. 3(a). The gap-like feature was already observed in STS on graphene [25]. It is explained by a suppression of electronic tunneling to graphene states near the Fermi energy with a large wave vector and a simultaneous enhancement of electronic tunneling at higher energy due to a phonon-mediated inelastic channel. The dip at +0.25 V corresponds to the energy position of the Dirac point. Subtracting the energy of the out-of-plane acoustic graphene phonon mode (67 mV), the Dirac point locates at 0.18 eV above the Fermi energy, indicating p-type doping in graphene with a hole concentration of $2 \times 10^{12}$ cm$^{-2}$. This is in good agreement with previous experimental and theoretical reports on similar samples [26].

The gap and the dip are not resolved in the wide-range spectra shown in Fig. 3(b) due to the larger tip-sample distance for this measurement (see the caption). However, the measured spectra taken on features A and B show an additional peak at +1.1 V and +1.4 V, respectively, which the flat area does not. This observation is statistically significant as shown in the histogram in Fig. SI.1(b). The peak energies for features A and B are qualitatively consistent with their contrast in the STM images. Assuming that the energy dependence of the density of states (DOS) of the tip and of the tunneling transmissivity are negligible, the tunnel current is proportional to the integral of the density of states of the sample from the Fermi energy to the bias voltage. In constant current mode, a feedback retracts the tip from the sample as the tunnel current increases. Therefore, the image contrast becomes brighter at the position where the integral density of states is larger. The integral of the STS in Fig. 3(b) from the Fermi energy to any voltage above the onset of peak A (+0.75 V) is larger for peak A than for peak B, since the peak energy of A is lower than that of B, and their peak intensities are almost the same. This is consistent with the fact that the STM contrast is brighter for features A than for features B for bias voltages above +0.75 V.

Inverse photoemission spectra showed that the buffer layer possesses a state at 1.1 eV above the Fermi energy, associated with Si dangling bonds, but it disappears after H intercalation.[27] STS on graphene on the C-face of SiC [SiC($000\bar{1}$)] also showed similar peaks.[28] This surface also has Si dangling bonds at the interface. At two spatial positions within a unit cell, hollow and top sites, where Si dangling bonds locate at the center of a graphene hexagon and on top of a graphene C atom, peak energies of +1.4 V and +1.6 V were obtained, respectively [28]. Here, for the case of QFMLG, the difference in the electronic states of the two types of features may also be attributed to different local graphene-Si stacking configurations.

The STS data can be interpreted based on the calculated DOS of the system with vacancies. In all examined systems, these display a common feature, i.e. a partially-filled peak localized across the Fermi level. We verified in all cases that this is an electronic state localized on the vacancy between the two layers, corresponding to a dangling bond (see the second row of Fig 4). For all vacancies with more than one missing H, a second peak appears which is not populated and whose energy increases with the size of the vacancy. It corresponds to an electronic state still localized at the vacancy, but with different symmetry, slightly more protruding upward and with electronic density interfering with that of the graphene π system. The total DOS for vacancies 1H, 3H, and 4H are reported in the SI. To compare the calculated DOS to the STS measurement, two effects must be considered. One is that we have approximated the local STS measurement with a calculated DOS evaluated in the whole cell. However, we underline that our calculation is performed on a supercell including only a few layers of substrate and whose lateral size is only slightly larger (at most double) than the extension of the vacancy itself. Therefore, our calculations already correspond to an "almost-local" measurement of similar resolution as the experimental one. In fact, local DOS calculations performed integrating on volumes smaller than the simulation cell show only slight increase of the relative weight of the localized state DOS with respect to that of graphene and the relative weight of the empty peak with respect to the filled one, and no improvement in the agreement with experimental data. However, this is not likely to be the main difference of the DOS with respect to the STS data. The already-discussed gap-like feature at the Fermi level implies a suppression and masking of the filled peak, which is consequently not visible in the STS measurement. We emulate this effect empirically by using a smoothened switch function suppressing the DOS around the Fermi level. The corrected DOS aligned to the Fermi levels is reported in Fig. 3(c). As can be seen, the 1H vacancy peak almost-completely disappears, while the 3H and 4H vacancies display unoccupied peaks at energies above the Fermi level. Their energy location is somewhat lower than in experiment, which is quite common in the evaluation of excited states in DFT. The relative location and displacement of the peaks is, however, reproduced, which strongly supports our assignment of peak A to the 3H vacancy and peak B to the 4H vacancy. We verified (data reported elsewere[24]) that the DOS is almost independent on the relative translation of the graphene lattice with respect to the vacancy. The DOS of 1H vacancies located on hollow and top positions of graphene lattice are basically indistinguishable, which also holds for the different possible 3H and 4H vacancies with different translations. Therefore, the A vs. B difference is not likely due to different vacancy location. This is not in contrast with the above-mentioned report for the C-face of SiC,[28] because there the Si dangling bond is much closer to the graphene sheet than for the case of QFMLG, which explains why in our case the exact location of the single vacant Si site is of minor importance. Note that even for the most inwardly bent case, the 4H vacancy, Si atoms do not form covalent bonds to graphene. Their dangling bonds act therefore as charged scattering centers for charge transport in QFMLG.

The complex voltage-dependence of the A- and B-type vacancies can be explained by the nature of these states. The STS peaks (corresponding to unpopulated states in the theoretical DOS) can be described as anti-bonding-like states whose electronic density interferes with the $\pi$ system and protrudes outwards with respect to the graphene sheet. Therefore, they are visible as bright features at the bias voltages corresponding to the peaks. This is supported by the theoretically emulated STM images reported in the fourth row of Fig. 4.

Since the peaks are located at different energies, the maximum brightness appears at different voltages, causing variable contrast. In theory and experiment the location of the peaks differ and one can not use the experimental voltage values for the evaluation of STM images. Therefore, STM images were evaluated at different voltages, in each case including the whole empty peak and excluding the filled one, which is not visible in the experiment. The brightness of the vacancies is similar, but as said, it is evaluated at different voltages. The 1H vacancy does not give appreciable contrast, because its DOS is located at the Fermi level, and therefore not visible in the STM experiment.

We additionally observe that the electronic state protrusion is somehow counterbalanced by the structural intrusion. While the latter is constant, the former is voltage dependent. Therefore, at a given voltage (away from the peak) the contrast can switch to negative. The change in contrast happens at different voltages for A and B features because the structural intrusion and the location/size of the peaks are different.

In summary, the phenomenology of AFM and STM measurements is consistent with the assignment of a larger vacancy to B and a smaller one to A. Specifically, 4H and 3H, respectively, are most consistent with the experimental data. Even in the most inward bent case, the 4H vacancy, the Si atoms do not form covalent bonds to graphene, but hold the dangling bonds which degrade the mobility of QFMLG. Furthermore, this study demonstrates that a combination of STM/AFM and DFT is the ideal tool to evaluate the distribution, size, and the electronic states of atomic-size defects. Using these techniques, we have probed the interactions between graphene and SiC. The atomic and electronic structures of charge scattering centers in QFMLG were for the first time experimentally and theoretically revealed in great detail. Our results provide insight into the relevant defects in this material system and might be valuable in determining the configuration of lowest free energy of QFMLG. This will provide guidelines for the optimization of the synthesis conditions of QFMLG such as substrate temperature, time, and pressure for the H intercalation process, in order to reduce the density of Si dangling bonds. Our work therefore provides the understanding for a rational design of high-mobility QFMLG. Besides, our approach in this study can be applied to investigate localized states in other systems, such as graphene interfaces intercalated with other elements.

We acknowledge travel support from COST Action MP1103 '*Nanostructured materials for solid-state hydrogen storage*'. Funding from the European Union Seventh Framework Program under Grant Agreement No. 696656 Graphene Flagship Core1 is also acknowledged. Financial support from the CNR in the framework of the agreements on scientific collaborations between CNR and CNRS (France), NRF (Korea), and RFBR (Russia) is acknowledged. We also thank the European Research Council (ERC) for funding under the European Union's Horizon 2020 research and innovation program (Grant Agreement No. 670173), the ERC Advanced Grant CEMAS (agreement no. 291194), the ERC Consolidator Grant AMSEL (682144), the EU project PAMS (610446), and the Initial Training Network QTea (317485), and Scuola Normale

Superiore for support via the internal project SNS16_B_HEUN – 004155. Furthermore, we acknowledge funding from the Italian Ministry of Foreign Affairs. We gratefully acknowledge CINECA for providing HPC resources under the ISCRA-C grants "Quasi free standing graphene monolayer on SiC with H-coverage vacancies: a density functional theory study" (2016-2017) and "Electro-mechanical manipulation of graphene" (2015-2016), and for technical support.

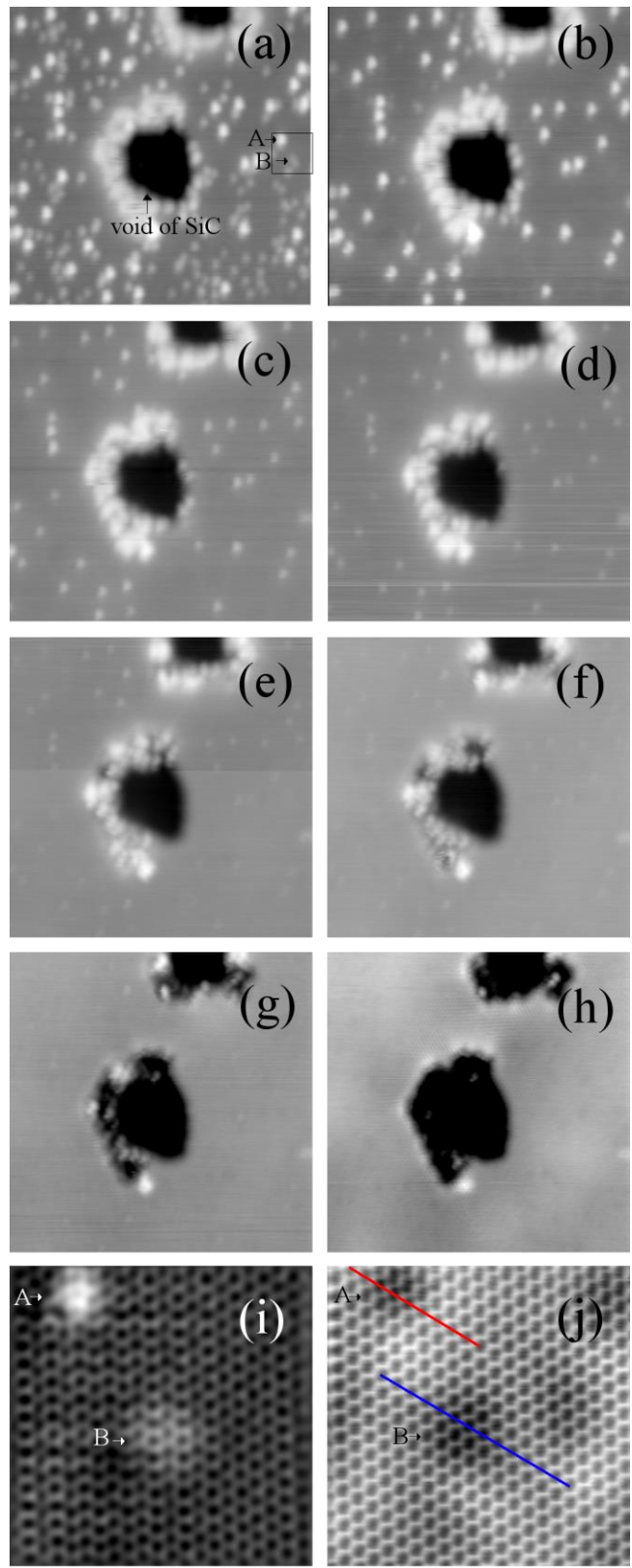

FIG. 1. (Color online) (a-i) STM images taken at bias voltages of (a) +1.5 V, (b) +1.0 V, (c) +0.75 V, (d) +0.5 V, (e) +0.3 V, (f) +0.2 V, (g) +0.1V, (h) -0.05 V, (i) +0.1 V and tunneling currents of (a-h) 5 pA and (i) 20 pA. (i) was taken in the area of the square in (a). (j) Constant-height AFM image taken in the same area as (i). The color scale ranges from -62.6 Hz (black) to -52.3 Hz (white). The image sizes are (a-h) 30 nm × 30 nm and (i, j) 4 nm × 4 nm. T = 5 K. Typical features A and B are indicated by arrows in (a), (i), and (j).

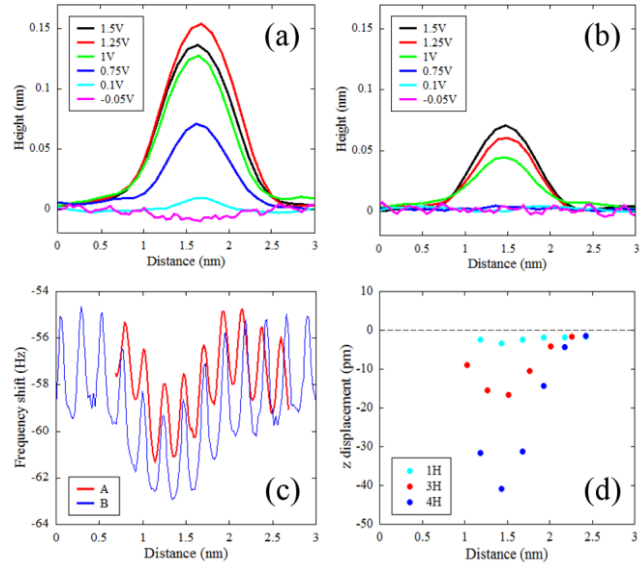

FIG. 2. (Color online) Line profiles of typical features (a) A and (b) B in STM images at various bias voltages and tunneling current of 5 pA. (c) Line profiles of the features A and B indicated by the red and blue lines along the [11$\bar{2}$0] direction in the AFM image shown in Fig. 1(f). T = 5 K. (d) DFT-calculated vertical displacement of C atoms along the [11$\bar{2}$0] direction for vacancies with 1 (cyan), 3 (red), and 4 (blue) missing H atoms.

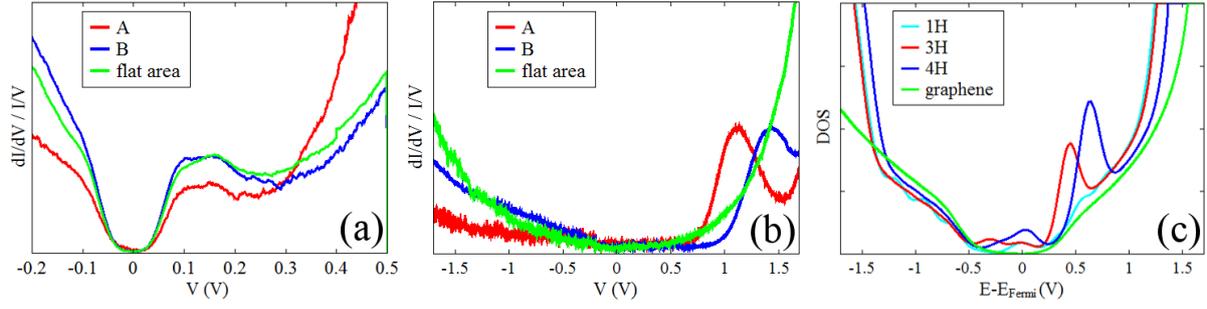

FIG. 3. (Color online) (a) and (b) STS on the features A and B and on a flat area. The energy range is (a) from -0.2 V to +0.5 V and (b) from -1.7 V to +1.7 V. T = 5 K. $V_{stab}$ is (a) 0.5 V and (b) 1.7 V, while $I_{stab}$ is 0.01nA for both. (c) Calculated DOS for vacancies 1H, 3H, 4H, and free graphene. For details, see text.

|  | 1H | 3H | 4H |
|---|---|---|---|
|  |  |  |  |
| Dangling bond |  |  |  |
| Simulated AFM |  |  |  |
| Simulated STM |  |  |  |

**Fig 4**: Theoretical analysis of the structures of top centered vacancies with 1, 3, and 4 missing H atoms (columns 1H, 3H, 4H), whose representation is reported in the first row (graphene lattice in red, Si in yellow, C of substrate in cyan, H in white, represented as enlarged spheres to better show the vacancy). The following properties are reported for the different kinds of vacancies in subsequent rows: an isosurface representation of the "dangling bond" state localized between graphene and SiC; the AFM-like images (obtained from the total electron density represented on a horizontal plane located between the tip and graphene, see the SI for details); the STM isocurrent-like images, taken at voltages including the whole empty peak and excluding the filled one. The STM images are obtained generating iso-surface representations of the state (iso-value = $10^{-6}$ e/Å$^3$) and color-coding it according to the height (for details, see

the SI). The periodicity is imposed by the model system and approximately corresponds to a completely occupied 6x6 superlattice.